\documentclass[11pt,a4paper]{article}
\usepackage{amsmath,amsthm}
\usepackage{graphicx}

\renewcommand{\thesection}{\Roman{section}}
\renewcommand{\thesubsection}{\Alph{subsection}}
\makeatletter
   \def\@seccntformat#1{\@ifundefined{#1@cntformat}%
   {\csname the#1\endcsname\quad}                  
   {\csname #1@cntformat\endcsname}}               
   \def\section@cntformat{\thesection.\hspace{0.4em}}
   \def\subsection@cntformat{\thesubsection.\hspace{0.4em}}
\renewcommand\@biblabel[1]{$^{#1}\!\!$}
\makeatother

\newcommand{\const}{\ensuremath{\mathrm{const}}}
\newcommand{\abs}[1]{\lvert#1\rvert}

\DeclareMathOperator{\erf}{erf}
\DeclareMathOperator{\sign}{sgn}

\newcommand{\Ndash}{\nobreakdash--}

\newcommand{\SBM}{\ensuremath{\mathit{SBM}^\pm}}
\newcommand{\ARN}{\ensuremath{\mathit{ARN}^\pm}}
\newcommand{\AF}{\ensuremath{\mathit{AF}^\pm}}
\newcommand{\OS}{\ensuremath{\mathit{OS}}}
\newcommand{\RN}{\ensuremath{\mathit{RN}}}
\newcommand{\RH}{\ensuremath{\mathit{RH}}}
\newcommand{\DegH}{\ensuremath{\mathit{DH}^\pm}}
\newcommand{\LW}{\ensuremath{\mathit{LW}}}
\newcommand{\LR}{\ensuremath{\mathit{LR}^\pm}}
\newcommand{\CR}{\ensuremath{\mathit{CR}}}

\theoremstyle{plain}
\newtheorem{Proposition}{Proposition}
\newtheorem*{Theorem}{Theorem}

\begin{document}
\vspace*{-2cm}
\begin{flushright}
     gr-qc/9906024
\end{flushright}

\bigskip \bigskip \bigskip

\begin{center}
 \textbf{\Large
     Dynamical system analysis \\
     for the Einstein--Yang--Mills equations} \\\bigskip
     M. Yu.\ Zotov$^*$  \\\medskip
     \textit{D. V. Skobeltsyn Institute of Nuclear Physics \\
     of Moscow State University, Moscow 119899, Russia} \\\bigskip
     \today
\end{center}

\bigskip
\bigskip
\centerline{\textbf{Abstract}}

\begin{quote}
     Local solutions of the static, spherically symmetric
     Einstein--Yang--Mills (EYM) equations with $SU(2)$ gauge group are
     studied on the basis of dynamical systems methods.  This approach
     enables us to classify EYM solutions in the origin neighborhood,
     to prove the existence of solutions with the oscillating metric as
     well as the existence of local solutions for all known formal
     power series expansions, to study the extendibility of solutions,
     and to find two new local singular solutions.

\medskip

     PACS: 05.45.-a, 04.20.Ex, 04.60.m
\end{quote}

\vfill
\rule{6cm}{0.4pt}

$^*$e-mail: \texttt{zotov@eas.npi.msu.su}


\newpage
\section{INTRODUCTION}

     The discovery of a discrete family of asymptotically flat
     particle-like solutions for the static, spherically symmetric
     Einstein--Yang--Mills (EYM) equations with $SU(2)$ gauge group,
     made by Bartnik and McKinnon in 1988~\cite{BMK}, evoked
     considerable interest in these equations and their various
     generalizations.  The intensity of investigations performed in
     this field is shown by the latest review~\cite{VGreview}, which
     summarizes a decade's work and contains more than three hundred
     references to publications on the subject.  However, there are
     still some problems which remain unsolved.  One of the most
     interesting of these is probably the task to prove the existence
     of the metric oscillations in the origin, $r=0$, neighborhood,
     which were found numerically during the study of the EYM black
     holes interior structure~\cite{grqc96}.  The initial purpose of
     the present work was to solve this problem.  Some other open
     questions can be found in~\cite{SW:extend}.

     Let us note that there are at least two approaches to the analysis
     of local solutions of nonlinear ordinary differential equations.
     One of them, namely, the asymptotic theory of differential
     equations, in some cases makes a possibility to obtain a complete
     classification of solutions in a singular point neighborhood.
     However, the right hand side of the studied equation must as a
     rule satisfy some rather specific conditions (see,
     e.g.,~\cite{KigCha}).  Another way, known as the theory of
     dynamical systems, or the qualitative theory of differential
     equations, is less restrictive in this sense, though it also does
     not always lead to a comprehensive description of the solutions
     behavior (see, e.g.,~\cite{Perko}).  Nevertheless, there are a
     number of problems in astrophysics and cosmology, which were
     solved basing on this approach (see~\cite{OB,DS4cosm} and
     references therein).

     In this paper, dynamical systems methods are used for the analysis
     of the EYM solutions asymptotic behavior.  This enables us to
     prove the existence of the above mentioned solutions with the
     oscillating metric,  as well as the existence of local solutions
     for all known formal power series expansions, and to find two new
     local solutions.  Moreover, a classification of local solutions in
     the vicinity of the origin is obtained.  In particular, it is
     shown that there exists a neighborhood of $r=0$ such that the
     metric function has a fixed sign in it.  Specifically, if the
     limiting value of the gauge function equals~$\pm 1$, then all real
     solutions belong to the Schwarzschild and Bartnik--McKinnon type.
     In other cases, the solution behavior depends on the metric
     function sign.  Namely, if the metric function is positive, then
     all solutions possess the behavior of the Reissner--Nordstr\"om
     type.  If, on the contrary, the metric function is negative, then
     almost all solutions are such that the metric function oscillates
     with its amplitude growing unboundedly as $r \to 0$, but the gauge
     function is monotonous (though its derivative also oscillates with
     the unboundedly growing amplitude).  Only particular solutions in
     this case exhibit asymptotic behavior of the
     ``anti--Reissner--Nordstr\"om'' type.  This result also gives the
     negative answer to the question stated in~\cite{SW:extend},
     whether $r=0$ is a limit point for zeros of the metric function.

     We have also considered the asymptotic behavior of solutions in
     the far field, $r \gg 1$, and in the vicinity of the points where
     the metric function tends to zero.  This analysis leads to a
     discovery of two new local singular solutions and allows us to
     obtain some conclusions concerning the extendibility of solutions
     and their limiting behavior as the number of the gauge function
     nodes tends to infinity.

     A detailed discussion of the physical interpretation of the
     EYM equations solutions can be found in~\cite{VGreview}.


\section{THE EQUATIONS}

     Recall that the space-time metric for the static, spherically
     symmetric EYM equations can be written as
\[
     ds^2 = \sigma^2 N \, dt^2 - N^{-1} \, dr^2
            - r^2 \left(d \vartheta^2
                    + \sin^2 \vartheta \, d \varphi^2 \right),
\]
     where $N$ and~$\sigma$ depend on~$r$, and the Yang--Mills
     gauge field reads as
\[
     A = (T_2 \, d \vartheta - T_1 \sin \vartheta \, d \varphi) \, w
         + T_3 \cos \vartheta \, d \varphi ,
\]
     where $T_i = \frac{1}{2} \, \tau _i$ are the~$SU(2)$ group
     generators and $\tau _i$ are the Pauli matrices, $i = 1,2,3$
     (see, e.g.,~\cite{VGreview}).

     The EYM equations in this framework take the form of two
     ordinary differential equations for the metric function~$N$
     and the gauge function~$w$:
\begin{equation} \label{eym4Nr}
 \begin{split}
     & r^3 N' + \bigl(1 + 2 {w'}^2 \bigr) r^2 N
              + \left(1 - w^2 \right)^2 - r^2 = 0, \\
     & r^3 N w'' - \left[\left(1 - w^2 \right)^2 - r^2
                 + r^2 N \right] w' + (1 - w^2) r w = 0,
 \end{split}
\end{equation}
     and a decoupled equation for~$\sigma$:
\begin{equation*}
     \frac{\sigma'}{\sigma} = \frac{2 {w'}^2}{r}.
\end{equation*}
     Since~\eqref{eym4Nr} do not involve~$\sigma$, one can use these
     to obtain~$N$ and~$w$, and then solve the equation for~$\sigma$.
     Thus we restrict our considerations to Eqs.~\eqref{eym4Nr}.  We
     also remark that~\eqref{eym4Nr} are invariant under the
     transformation $r \to -r$; thus, in what follows we discuss only
     the region $r \ge 0$.

     For the purposes of studying the EYM solutions at finite~$r$,
     it is convincing to rewrite~\eqref{eym4Nr} in terms of~$w$
     and $u = r^2 N$.  They become
\begin{equation} \label{EYM}
 \begin{split}
     & r u'    - \bigl(1 - 2 {w'}^2 \bigr) u
               + \left(1 - w^2 \right)^2 - r^2 = 0, \\[2pt]
     & r u w'' - \left[u + \left(1 - w^2 \right)^2 - r^2 \right] w'
               + \left(1 - w^2 \right) r w = 0.
 \end{split}
\end{equation}

     Recall that the only known explicit solutions of~\eqref{EYM}
     are the Schwarzschild solution
\begin{equation} \label{ExactSchwarz}
     w \equiv \pm 1,                     \quad
     u = a \, r + r^2,
\end{equation}
     and the Reissner--Nordstr\"om solution
\begin{equation} \label{ExactRN}
     w \equiv 0,                          \quad
     u = 1 + b \, r + r^2,
\end{equation}
     where $a$ and $b$ are arbitrary constants.

     In order to apply the theory of dynamical systems to the analysis
     of the EYM equations~\eqref{EYM}, it is necessary to write them as
     an autonomous system of first-order differential equations.  Thus
     we introduce the function $v = w'$ and an independent variable~$t$
     defined by $dr = r u \, dt$. After making these changes, we obtain
     the dynamical system
\begin{equation} \label{DS}
 \begin{split}
     \dot{r} & = r u, \\
     \dot{u} & = \left[\left(1 - 2 v^2 \right) u
                 - \left(1 - w^2 \right)^2 + r^2 \right] u, \\
     \dot{v} & = \left[ u + \left(1 - w^2 \right)^2 - r^2 \right] v
                 - \left(1 - w^2 \right) r w, \\
     \dot{w} & = r u v.
 \end{split}
\end{equation}
     Notice that this system has solutions for $r \equiv 0$ and
     $u \equiv 0$, which do not take place for~\eqref{EYM}.

     The first step to start analyzing~\eqref{DS} is to determine the
     critical points.  It is easy to verify that
     the dynamical system~\eqref{DS} has the following critical
     sets:
\[
 \begin{array}{ll}
     \ARN  \colon & (0, -(1 - w^2)^2, \pm 1, w), \\
     \RN   \colon & (0, (1 - w^2)^2, 0, w), \\
     \SBM  \colon & (0, 0, v, \pm 1), \\
     W     \colon & (0, 0, 0, w), \\
     \RH   \colon & (r, 0, (1 - w^2)r w/[(1 - w^2)^2 - r^2], w), \\
     \DegH \colon & (\pm 1, 0, v, 0).
 \end{array}
\]
     All the critical sets belong to the hyperplanes $r=0$ and/or
     $u=0$, in which the conditions of the existence--uniqueness
     theorem do not hold for~\eqref{EYM}.

     In what follows we shall not give a global
     phase portrait for~\eqref{DS}, but we shall mainly concentrate
     on the results that have direct consequence for the EYM
     equations~\eqref{EYM}.


\section{PRELIMINARY INVESTIGATION OF THE \\ ORIGIN NEIGHBORHOOD}

     Let us consider the projection of~\eqref{DS} into the hyperplane
     $r = 0$.  This immediately leads to $w \equiv w_0 = \const$.  Thus
     the dynamical system~\eqref{DS} reduces to
\begin{equation} \label{ds4uv}
 \begin{split}
     \dot{u} & = -\alpha^2 u + \left(1 - 2 v^2 \right) u^2, \\
     \dot{v} & = \alpha^2 v + u v,
 \end{split}
\end{equation}
     where $\alpha = 1 - w_0^2$.  Notice that~\eqref{ds4uv}
     is invariant under the transformation $v \to -v$.  Hence,
     the phase portrait will be symmetric with respect to the
     $u$-axis.

     Since~\eqref{ds4uv} contains a free parameter~$\alpha$,
     it is convincing to split the analysis into two steps.
     Let us begin with $\alpha=0$.  In this case,~\eqref{ds4uv}
     reads as
\begin{equation} \label{uv4wpm1}
 \begin{split}
     & \dot{u} = \left(1 - 2 v^2 \right) u^2, \\
     & \dot{v} = u v.
 \end{split}
\end{equation}
     One can easily solve this system.  First, the critical points,
     which are the projection of the sets~\SBM{} and~$W$, give
     $u \equiv 0$, $v \equiv \const$.  Next,
\[
     C_2 - C_1 t  =     \frac{1}{v} \, e^{v^2}
     - 2 \int _0^v e^{\tilde{v}^2} \, d \tilde{v}=
     \frac{1}{v} \, e^{v^2} + i \sqrt{\pi} \, \erf (i v),
     \quad \text{and} \quad u = C_1 v \, e^{-v^2},
\]
     where $C_1$ and $C_2$ are arbitrary constants, and $v \ne 0$.
     Finally, for $v \equiv 0$ one has  $u = (C - t)^{-1}$,
     where $C$ is an arbitrary constant,  $t \ne C$.
     We remark that if $u < 0$, then the nontrivial solutions tend
     to zero as $t \to +\infty$.
     In the opposite case, they tend to zero as
     $t \to -\infty$.  In particular, it follows that
     if $\lim_{r \to 0} w(r) = \pm 1$ for a solution
     of~\eqref{EYM}, then $\lim_{r \to 0} w'(r) = 0$.

\begin{figure}[t]
\begin{center}
 \includegraphics[width=0.7\textwidth]{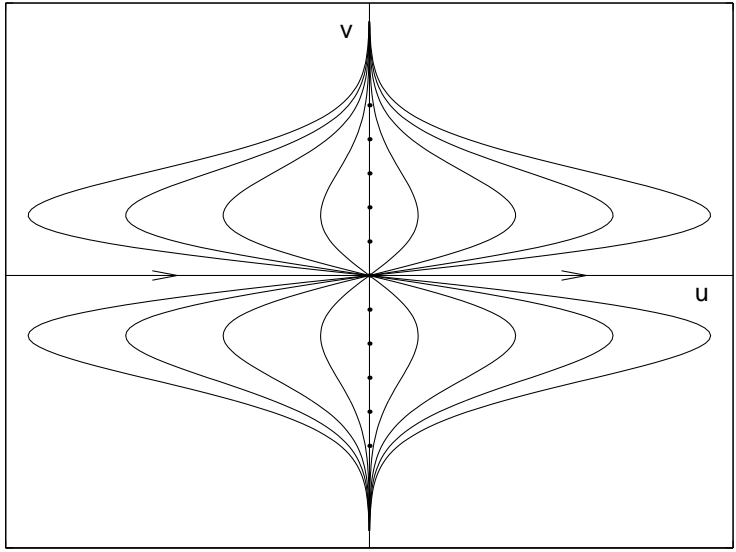}
\end{center}
 \caption{Phase portrait of the dynamical system~\eqref{uv4wpm1}.}
 \label{PhP4wpm1}
\end{figure}

     The phase portrait of~\eqref{uv4wpm1} is shown in
     Fig.~\ref{PhP4wpm1}.

\medskip

     Now let us study~\eqref{ds4uv} for $\alpha \ne 0$ (i.e.,
     for $w_0 \ne \pm 1$).  In this case, the system~\eqref{ds4uv}
     has the following critical points:  $Z$: $(0,0)$,
     $A^\pm$: $(-\alpha^2, \pm 1)$, and $R$: $(\alpha^2, 0)$.

\begin{figure}[t]
 \begin{center}
 \includegraphics[width=0.75\textwidth]{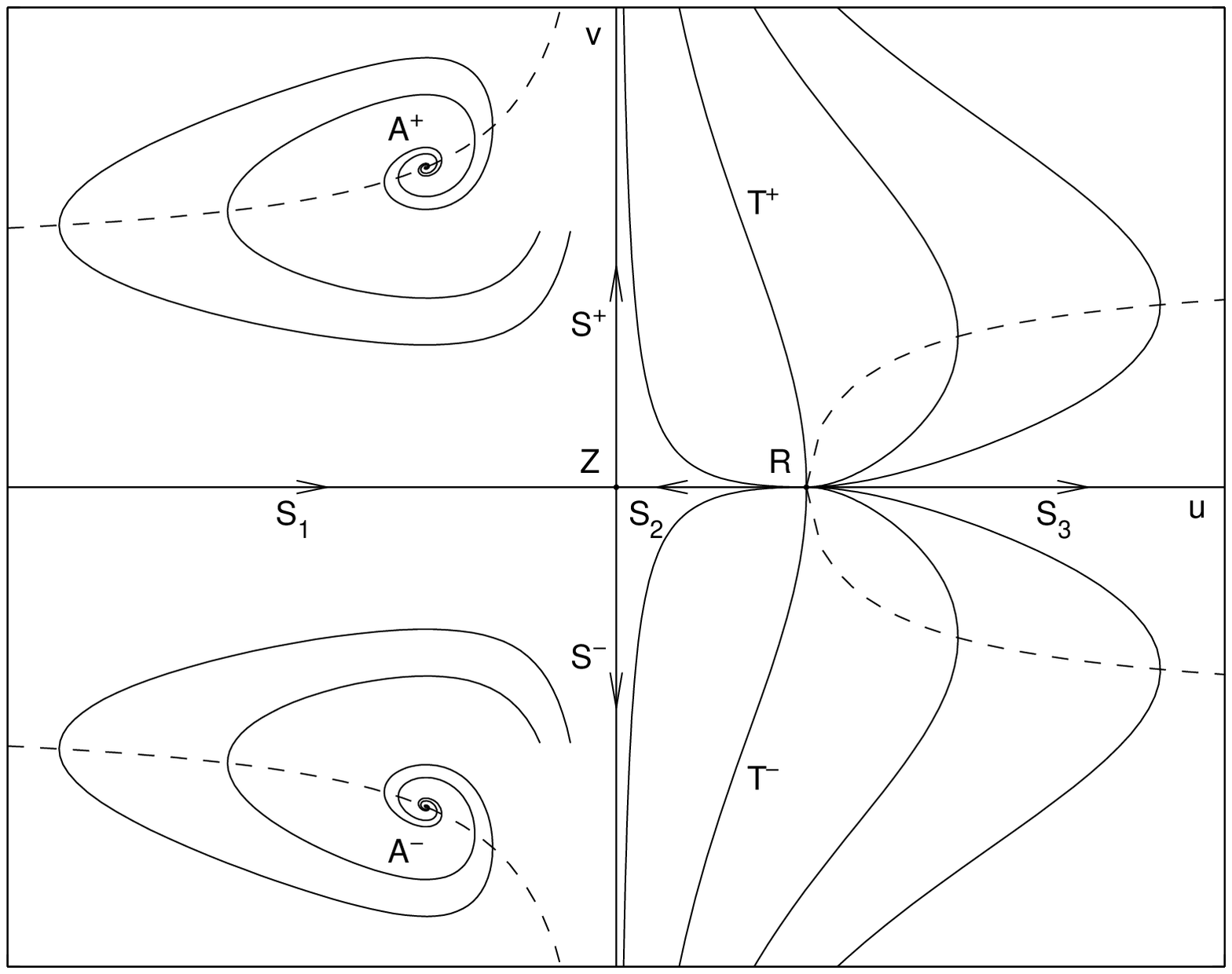}
 \end{center}
 \caption{Phase portrait of the dynamical system~\eqref{ds4uv}
in the vicinity of the finite critical points for $\alpha \ne 0$.
Dashed lines represent the curves $u'(v) = 0$.}
 \label{PhP4uv}
\end{figure}

     The point $Z$ is the projection of~$W$.
     The eigenvalues of~$Z$ are $\lambda_u = -\alpha^2$
     and $\lambda_v = \alpha^2$. Thus, $Z$~is a saddle.
     It has four separatrices, which can be easily obtained
     explicitly.  The repelling separatrices, denote them by~$S^\pm$
     in accordance with the sign of~$v$,
     are tangent to the eigenvector $\zeta_v = (0,1)$.
     They  can be written as
\begin{equation} \label{sepSpm}
             u \equiv 0,                                    \quad
             v = C_v \, e^{\alpha^2 t}.
\end{equation}
     Here and forth the letter $C$, with an alphabetical
     subscript ($C_u$, $C_v$, etc.), denotes a nonzero constant.

     The attracting separatrices are tangent to the eigenvector
     $\zeta_u = (1,0)$.  They take the form
\begin{equation} \label{sepS123}
     u = \frac{\alpha^2}{1 + C_u \, e^{\alpha^2 t}},        \quad
     v \equiv 0.
\end{equation}
     One of these separatrices, denote it by~$S_1$, belongs to
     the half-plane $u < 0$. For~$S_1$,
     $C_u < 0$ and $t > - \alpha^{-2} \ln \abs{C_u}$.
     Another separatrix, $S_2$, lies in the half-plane $u > 0$.
     It has $C_u > 0$ and joins~$Z$ to~$R$ (see Fig.~\ref{PhP4uv}).

     The point $R$ is the projection of the critical curve~\RN.
     The eigenvalues of $R$ are $\lambda_u = \alpha^2$ and
     $\lambda_v = 2 \alpha^2$.  Thus, it is an unstable node.
     Almost all trajectories that approach~$R$ as $t \to -\infty$,
     are tangent to the eigenvector $\zeta_u = (1,0)$.
     The corresponding separatrices have the form~\eqref{sepS123}.
     One of them, namely, $S_2$ joins~$R$ to~$Z$.
     Another one, $S_3$, is defined for $C_u < 0$ and
     $t < - \alpha^{-2} \ln \abs{C_u}$.

     There are also two separatrices, denote them by~$T^\pm$,
     which are tangent to the eigenvector $\zeta_v = (0, 1)$.
     Let us show that they have the form
\begin{equation} \label{ExclSol}
     u = \alpha^2 \left( 1 - \tfrac{2}{3} v^2 \right) + o(v^2)
\end{equation}
     as  $v \to 0$.
     To see this, define $u_v = (u - \alpha^2) v^{-1}$.  Now the
     dynamical system~\eqref{ds4uv} reads as
\[
 \begin{split}
     \dot{u}_v & = - \alpha^2 \left(u_v + 2 \alpha^2 v \right)
                   - 2 \left(2 \alpha^2 + u_v v \right) u_v v^2, \\
     \dot{v}   & = 2 \alpha^2 v + u_v v^2.
 \end{split}
\]
     For $v = 0$, this system has a saddle $(0,0)$ with
     the eigenvalues $\lambda _{u_v} = - \alpha^2$,
     $\lambda _v     = 2 \alpha^2$ and the eigenvectors
     $\zeta _{u_v} = (1,0)$,
     $\zeta _v     = \left(-\tfrac{2}{3} \alpha^2, 1 \right)$.
     The separatrices that are tangent to the eigenvector~$\zeta_{u_v}$,
     belong to the line $v = 0$.  Hence, there are no corresponding
     trajectories of~\eqref{ds4uv}. Conversely, the eigenvector~$\zeta_v$
     determines the outgoing separatrices, which take the form
     $u_v = -\tfrac{2}{3} \alpha^2 v + o(v)$ as  $v \to 0$.
     This yields~\eqref{ExclSol}.

     Note that the same technique can be used to find the higher order
     terms in~\eqref{ExclSol}.  This is also valid for the asymptotic
     solutions presented below.

\medskip

     Finally, the points $A^\pm$ represent projections of
     the critical curves~\ARN.  The eigenvalues of~$A^\pm$
     are $\lambda_{u,v} = \frac{1}{2} \alpha^2 (1 \pm i \sqrt{15})$.
     Thus, these critical points are repelling foci.
     It is important to note here that for all the trajectories
     that spiral away from~$A^\pm$, the metric function~$u$ is
     strictly negative because of the separatrices~$S^\pm$,
     and~$v$ preserves its sign due to the separatrix~$S_1$
     (or, the same, because of the above mentioned invariance
     of~\eqref{ds4uv} under the transformation $v \to -v$).
     Since there are no other finite critical points in the
     half-plane $u < 0$, the trajectories that spiral away
     from the points~$A^\pm$, do not have limit cycles.  We remark
     that this can also be easily proved if we define the Dulac
     function for~\eqref{ds4uv} by $(u^2 v)^{-1}$.
     Hence,~$u$ and~$v$ exhibit oscillations with the amplitude
     growing infinitely as $t \to +\infty$.

     Fig.~\ref{PhP4uv} shows the phase portrait of~\eqref{ds4uv} near
     the points~$A^\pm$, $Z$, and~$R$.  Notice that this portrait is
     drastically different from that one shown in Fig.~\ref{PhP4wpm1}.
     Thus, $\alpha$ is a bifurcation parameter for the dynamical
     system~\eqref{ds4uv}.


     In order to obtain the global phase portrait of~\eqref{ds4uv},
     one has to study the behavior of its trajectories at infinity.
     Using the standard transform to the projective coordinates,
     one can find out that the system~\eqref{ds4uv} has
     four critical points at the $(u,v)$ phase plane boundary,
     namely, $U^\pm$: $(u = \pm \infty, v = 0)$ and
     $V^\pm$: $(u = 0, v = \pm \infty)$.  The points~$U^\pm$ are
     saddles, and~$V^\pm$ are saddle-nodes.  The separatrices~$S_3$
     and~$S_1$ are the only trajectories, which approach the
     points~$U^\pm$ from the finite region of the phase plane.  The
     points~$V^\pm$, besides the separatrices~$S^\pm$, have ingoing
     trajectories, which emanate from~$R$.  These trajectories have
     the form $u = \tfrac{1}{2} \alpha^2 \, v^{-2} + o(v^{-2})$
     as  $v \to \infty$.
     The boundary of the phase plane contains two separatrices
     that join~$U^+$ to~$V^\pm$ and two separatrices that
     go from~$V^\pm$ to~$U^-$.


\section{THE ORIGIN NEIGHBORHOOD} \label{section:r=0}


\subsection{The critical curves \ARN}

     Let us turn to the analysis of the dynamical system~\eqref{DS}
     near the critical curves \ARN: $(0, -(1 - w^2)^2, \pm 1, w)$
     for $w \ne \pm 1$.  The excluded points also belong to the
     lines~\SBM{} and will be studied below.  Notice that the
     curves~\ARN{} (with the points $w = \pm 1$ excluded) lie
     in the region $u < 0$.  Hence, in a neighborhood of these
     curves, $t$ growing to infinity corresponds to decreasing~$r$
     in the EYM equations~\eqref{EYM}.

     The eigenvalues of \ARN{} are $\lambda _r     = - (1 - w^2)^2$,
     $\lambda _{u,v} = \tfrac{1}{2} (1 - w^2)^2 (1 \pm i \sqrt{15})$,
     and $\lambda _w     = 0$.
     It should be recalled at this point that an
     $n$\nobreakdash-\hspace{0pt}dimensional critical set
     necessarily has~$n$ zero eigenvalues (see, e.g.,~\cite{OB}).
     Thus, the zero eigenvalue~$\lambda_w$ corresponds to the fact
     that~\ARN{} are one-dimensional sets of critical points.
     Since other eigenvalues have nonzero real part, \ARN{} are
     hyperbolic sets.

     The eigenvalues~$\lambda_{u,v}$ determine three-dimensional
     unstable manifolds~$M^\pm$ of the curves~\ARN{}, respectively.
     Since~$\lambda_{u,v}$ are complex, the trajectories that lie
     on~$M^\pm$, describe oscillatory behavior of~$u$ and~$v$.
     The found above trajectories that spiral away from the
     points~$A^\pm$, are the projection of the trajectories that
     lie on~$M^\pm$, in the plane $(r = 0, w = w_0 \ne \pm 1)$.
     The separatrices~$S^\pm$ and~$S_1$ do also have the
     obvious counterparts for~\eqref{DS}:
\begin{equation} \label{FullSep4u}
     r \equiv 0,                        \quad
     u \equiv 0,                        \quad
     v = C_v \, e^{\alpha^2 t},         \quad
     w \equiv w_0,
\end{equation}
     and
\begin{equation} \label{FullSep4v}
     r \equiv 0,                                     \quad
     u = \frac{\alpha^2}{1 + C_u \, e^{\alpha^2 t}}, \quad
     v \equiv 0,                                     \quad
     w \equiv w_0,
\end{equation}
     respectively, where $w_0 = \const \ne \pm 1$, and $C_u < 0$.
     These two-dimensional separatrices preserve the signs of~$u$
     and~$v$ for the trajectories on~$M^\pm$.

     Recall that the trajectories that spiral away from~$A^\pm$
     do not have limit cycles.  Evidently, the same is valid
     for the trajectories on~$M^\pm$.

     Next, due to the negative eigenvalue~$\lambda_r$, each
     of the curves~\ARN{} has two-dimensional stable separatrices,
     which are tangent to the eigenvectors
\[
     \zeta _r^\pm = \left(1, \pm 4 (1 - w^2) w,
                          \frac{w}{1 - w^2}, \pm 1 \right),
\]
     where the upper sign applies for~$\mathit{ARN}^+$ and the lower
     one for~$\mathit{ARN}^-$.  It is easy to see that these
     separatrices correspond to a one-parameter family of the EYM
     solutions that exist in the origin neighborhood.
     Thus, we conclude with

\begin{Proposition}
     Let $\mathcal{U}^-$ be a set of solutions for the EYM
     equations~\eqref{EYM} such that they are defined in some
     neighborhood of $r = 0$, $u(r) < 0$ in this neighborhood,
     and $\lim _{r \to 0} w(r) = w_0 < \infty$,  $w_0 \ne \pm 1$.
     Then $\mathcal{U}^-$ is nonempty.  Moreover, almost
     all solutions of~\eqref{EYM} that belong to~$\mathcal{U}^-$,
     are monotonous for the gauge function~$w$ and oscillating
     for the metric function~$u$.  These solutions have the following
     properties:
\begin{enumerate}
 \item
     The amplitude of the metric function oscillations grows unboundedly
     as $r \to 0$.
 \item
     The values of the metric function at the points of maximum form
     a sequence, which monotonically converges to zero as $r \to 0$.
 \item
     The derivative of the gauge function also oscillates
     with the amplitude growing unboundedly as $r \to 0$,
     and gets closer to zero on each cycle of the oscillations,
     but its sign remains unchanged.
\end{enumerate}
     Besides these solutions, $\mathcal{U}^-$ also contains
     a one-parameter family of local solutions of the
     ``anti--\hspace{0pt}Reissner--\hspace{0pt}Nordstr\"om'' type:
\begin{equation} \label{ARN}
 \begin{split}
     u & = - (1 - w_0^2)^2 \pm 4 (1 - w_0^2) w_0 r + o(r), \\
     w & = w_0 \pm r + \frac{w_0}{2 (1 - w_0^2)} \, r^2 + o(r^2)
 \end{split}
\end{equation}
     as  $r \to 0$, where~$w_0 \ne \pm1$.
\end{Proposition}

     Formal expansions of the form~\eqref{ARN} and some
     corresponding numerical solutions were found
     in~\cite{grqc96}.  This paper was also the first to
     present the oscillating solutions.  Some of their
     properties were analyzed in~\cite{JETP}.


\subsection{The critical curve \RN}

     Now let us study~\eqref{DS} in the vicinity of the
     curve~\RN: $(0, (1 - w^2)^2, 0, w)$ for $w \ne \pm 1$.
     Similar to the above, the excluded points also belong to
     the critical sets~\SBM{} (and~$W$) and will be studied
     below.  Notice that~\RN{} (with the points $w =\pm 1$
     excluded) lies in the region $u > 0$.  Hence, in
     a neighborhood of~\RN, $t$ growing to infinity corresponds
     to increasing~$r$ in the EYM equations~\eqref{EYM}.

     The eigenvalues of \RN{} are
     $\lambda_r = \lambda_u = (1 - w^2)^2$,
     $\lambda_v = 2 (1 - w^2)^2$, and $\lambda_w = 0$.
     Therefore, all trajectories of~\eqref{DS} in the vicinity
     of~\RN{} belong to an unstable four-dimensional manifold;
     they correspond to a three-parameter family of EYM solutions.

     The eigenvalue~$\lambda_v$ determines two-dimensional
     separatrices, which are tangent to the eigenvector
     $\zeta_v = (0,0,1,0)$ and take the form
\[
     r \equiv 0,                                                \quad
     w \equiv \const \ne \pm 1,                                 \quad
     u = \alpha^2 \left(1 - \tfrac{2}{3} v^2 \right) + o(v^2)   \quad
     \text{as } v \to 0.
\]
     The separatrices~$T^\pm$ found above, represent their projection
     in the plane $(r=0, w=w_0 \ne \pm 1)$. The EYM equations~\eqref{EYM}
     do not have any corresponding solution. Conversely, the eigenvectors
     $\zeta_r = (1, 0, w/(1 - w^2), 0)$ and $\zeta_u = (0,1,0,0)$
     determine trajectories, which correspond to the EYM solutions
\begin{equation} \label{RN}
 \begin{split}
     u & = (1 - w_0^2)^2 + u_1 r + o(r), \\
     w & = w_0  + \frac{w_0}{2 (1 - w_0^2)} \, r^2 + o(r^2)
 \end{split}
\end{equation}
     as $r \to 0$, where $w_0 \ne \pm 1$ and $u_1$ are arbitrary
     constants, and the higher order terms contain one more parameter.

     Let us show how one can choose the third parameter
     in~\eqref{RN}.  The procedure will be similar to that one
     used for obtaining~\eqref{ExclSol}. Namely, consider~\eqref{DS}
     in the local coordinates
\[
     r,                                                     \quad
     u_r = \frac{u - (1 - w^2)^2}{r},                       \quad
     v_r = \frac{v}{r},                                     \quad
     w.
\]
     Then the corresponding dynamical system, which we omit for
     brevity, has the critical surface $(0, u_r, w/(1-w^2), w)$.
     (The other critical sets either have $w = \pm 1$ or do not
     belong to the hyperplane $r = 0$.)  The eigenvalues of this
     surface are $\lambda _r = \lambda _{v_r} = (1 - w^2)^2$ and
     $\lambda _{u_r} = \lambda _w = 0$.  It follows that all
     trajectories in a neighborhood of this surface belong to an
     unstable four-dimensional manifold.  The nonzero eigenvalues
     have the eigenvectors $\zeta_r = (1, 1 + 2 w^2, 0, 0)$ and
     $\zeta_{v_r} = (0,0,1,0)$. Thus, all the trajectories assume
     the form
\[
     u_r = u_1 + (1 + 2 w_0^2) \, r + o(r),                 \quad
     v_r = \frac{w_0}{1-w_0^2} + v_1 r + o(r),              \quad
     w   = w_0 + o(r)
\]
     as $r \to 0$, where $w_0 \ne \pm 1$, $u_1$, and $v_1$ are
     arbitrary constants.   Changing back to the initial variables
     and taking into account the above discussion, one has

\begin{Proposition}
     Let $\mathcal{U}^+$ be a set of solutions for the EYM
     equations~\eqref{EYM} such that they are defined in some
     neighborhood of $r = 0$, $u(r) > 0$ in this neighborhood,
     and $\lim _{r \to 0} w(r) = w_0 < \infty$,  $w_0 \ne \pm 1$.
     Then $\mathcal{U}^+$ is nonempty.  Moreover, all solutions
     of~\eqref{EYM} that belong to~$\mathcal{U}^+$, form
     a three-parameter family of local solutions of
     the Reissner--\hspace{0pt}Nordstr\"om type:
\begin{equation} \label{RN3}
 \begin{split}
     u & = (1 - w_0^2)^2 + u_1 r + r^2 + o(r^2), \\
     w & = w_0  + \frac{w_0}{2 (1 - w_0^2)} \, r^2 + w_3 r^3 + o(r^3)
 \end{split}
\end{equation}
     as $r \to 0$, where $w_0$, $u_1$ and $w_3$ are arbitrary
     constants, $w_0 \ne \pm 1$.
\end{Proposition}

     A formal power series expansion~\eqref{RN3} was presented
     in~\cite{VG90}.  Some black hole solutions with this asymptotic
     were first found numerically in~\cite{grqc96}.  The local
     existence proof for these solutions was given in~\cite{SW:RNlike}.
     We remark that here we follow the terminology, introduced
     in~\cite{grqc96}, which is slightly different from that one used
     in~\cite{SW:RNlike} and~\cite{SW:extend}.


\subsection{The critical lines \SBM}

     The lines \SBM: $(0,0,v,\pm 1)$ are degenerate critical sets,
     since the eigenvalues~$\lambda_r$, $\lambda_u$, and $\lambda_w$
     are equal to zero.  In order to study the behavior of trajectories
     of~\eqref{DS} in a neighborhood of these lines, we use the
     standard technique~\cite{OB}.

     First, define~$\bar{w}$ by $w = \bar{w} \pm 1$, where the
     upper sign applies for~$\mathit{SBM}^+$ and the lower one
     for~$\mathit{SBM}^-$.  Now the lines~\SBM{} are transformed
     to the $v$-axis.  Next, introduce the local coordinates
\begin{equation} \label{Mu4SBM}
     r_u = \frac{r}{u},                                \quad
     u,                                                \quad
     v,                                                \quad
     w_u = \frac{\bar{w}}{u},
\end{equation}
     in which the dynamical system~\eqref{DS} reads as
\begin{equation} \label{ds4Mu}
 \begin{split}
     \dot{r}_u & = (2 v^2 + K) \, r_u, \\
     \dot{u}   & = (1 - 2 v^2 - K) \, u, \\
     \dot{v}   & = (1 + K) \, v
                   + (1 \pm u w_u) (2 \pm u w_u) r_u u w_u, \\
     \dot{w}_u & = - (1 - 2 v^2 - K) \, w_u + r_u v, \\
 \end{split}
\end{equation}
     where $K = [(2 \pm u w_u)^2 w_u^2 - r_u^2 ] u$, and an overdot
     stands for derivatives with respect to~$\tau$ defined
     by $d\tau = u \, dt$ (thus, $\tau = \ln r + \const$).

     The system~\eqref{ds4Mu} has one critical set in the hyperplane
     $u = 0$, namely, the $r_u$-axis, which is an unstable hyperbolic
     line.  The corresponding eigenvalues are
     $\lambda _{r_u} = 0$,
     $\lambda _{u}   = \lambda _{v} = 1$, and
     $\lambda _{w_u} = - 1$.
     The two-dimensional ingoing separatrices
\[
     r_u \equiv \const{},                              \quad
     u   \equiv 0,                                     \quad
     v   \equiv 0,                                     \quad
     w_u =      C_w \, e^{-\tau},
\]
     which are tangent to the eigenvector $\zeta_{w_u} = (0,0,0,1)$,
     belong to the hyperplane $u = 0$.  Hence, they do not correspond
     to any trajectories of~\eqref{DS}.  In their turn, the
     eigenvalues~$\lambda_{u}$ and~$\lambda_{v}$, which have
     the eigenvectors $\zeta _u = (-r_u^3,1,0,0)$ and
     $\zeta _v = (0,0,2,r_u)$, determine the outgoing three-dimensional
     separatrices
\[
     r_u = r_0 - r_0^3 \, u + o(u),                    \quad
     v   = 2 v_1 \, u + o(u),                          \quad
     w_u = r_0 v_1 \, u + o(u)
\]
     as $u \to 0$,
     where $r_0$ and $v_1$ are arbitrary constants.  This implies

\begin{Proposition} \label{prop4SBM}
     All solutions of the EYM equations~\eqref{EYM} such that
     $\lim_{r \to 0} w(r) = \pm 1$, belong to a two-parameter family
     of local solutions of the Schwarzschild type:
\begin{equation} \label{as4SBM}
 \begin{split}
     u & = u_1 r + r^2 + o(r^2), \\
     w & = \pm 1 + w_2 r^2 + o(r^2)
 \end{split}
\end{equation}
     as $r \to 0$,
     where $u_1$ and~$w_2$ are arbitrary constants.
\end{Proposition}

     In particular, these local solutions describe the
     behavior of the Bartnik--McKinnon particle-like solutions~\cite{BMK}
     (for $u_1 = 0$) and the black hole solutions of the Schwarzschild
     type~\cite{grqc96} in the vicinity of the origin.

     It is interesting to note that the separatrices that are tangent
     to the eigenvector~$\zeta_u$, can be written as
\[
     u   =      \frac{1 + u_1 r_u}{r_u^2},             \quad
     v   \equiv 0,                                     \quad
     w_u \equiv 0,
\]
     where $u_1$ is an arbitrary constant, and $r_u \ne 0$.  Obviously,
     these separatrices correspond to the Schwarzschild
     solution~\eqref{ExactSchwarz}.

     We also remark that the coordinates~\eqref{Mu4SBM} enable us to
     resolve the degeneracy of the $v$-axis along the $u$-direction.
     Analysis of the $r$- and $w$-directions leads to the same
     conclusion for the EYM equations as stated in
     Proposition~\ref{prop4SBM}.

     It is necessary to underline here that we discuss only real
     EYM solutions, though the EYM equations also possess
     complex solutions.  For example, a study of~\eqref{DS}
     in the local coordinates
 $(r, u/r^2, v, \bar{w}/r)$,
     leads to a discovery of complex EYM solutions of the form
\[
 \begin{split}
     & u = - \left(1 + 4 w_1^2 \right)
             \left(r^2 + w_0 w_1 r^3 \right) + o(r^3), \\\smallskip
     & w = w_0 + w_1 r - \tfrac{1}{8} w_0 r^2 + o(r^2)
 \end{split}
\]
     as $r \to 0$,
     where $w_0 = \pm 1$ and $w_1 = \pm (\sqrt{3} \pm i \sqrt{5})/4$.
     These solutions do not have free parameters.


\subsection{The critical line $W$}

     The eigenvalues of the critical line~$W$: $(0,0,0,w)$ are
     $\lambda _r = 0$, $\lambda _u = - (1 - w^2)^2$,
     $\lambda _v = (1 - w^2)^2$, and  $\lambda _w = 0$.
     Hence, $W$ is degenerate for any~$w$.  The eigenvalues~$\lambda_u$
     and~$\lambda_v$ are nonzero and have different signs whenever
     $w \ne \pm 1$.  In this case, $W$ is an unstable critical set
     with the outgoing two-dimensional separatrices~\eqref{FullSep4u}
     and the ingoing two-dimensional separatrices~\eqref{FullSep4v}.
     Investigation of~\eqref{DS} in the vicinity of~$W$ gives the
     same result for the EYM equations as already stated in
     Proposition~\ref{prop4SBM}. By this reason we omit the discussion.


\medskip

     Thus, we have obtained a description of the EYM solutions that
     have finite values of the gauge function~$w$ in the origin
     neighborhood.  It follows from our considerations that for all
     these solutions $\lim _{r \to 0} w(r) = w_0 < \infty$.
     A standard analysis
     of the dynamical system~\eqref{DS} in the projective coordinates
     $(rz, uz, vz, z = 1/w)$ reveals that Eqs.~\eqref{EYM} do not
     have solutions such that $\lim _{r \to 0} w(r) = \infty$.  Hence,
     the results of this section can be summarized
     in the following classification of the EYM solutions, defined in
     the vicinity of the origin.

\begin{Theorem}
     All real solutions of the EYM equations~\eqref{EYM},
     defined in a neighborhood of $r = 0$, belong to one of
     the following disjoint classes:
\begin{enumerate}
 \item
     $w_0 = \pm 1$.  In this case, all solutions are of
     the Schwarzschild and Bartnik--\hspace{0pt}McKinnon
     type~\eqref{as4SBM}.

 \item
     $w_0 \ne \pm 1$,
     and the metric function~$u$ is negative in some neighborhood
     of $r = 0$.  In this case, almost all solutions are such that
     the metric function oscillates with the unboundedly  growing
     amplitude
     as $r \to 0$, but the gauge function is  monotonous (though its
     derivative also oscillates with the amplitude growing infinitely).
     Only particular solutions in this case exhibit asymptotic
     behavior of the ``anti--Reissner--Nordstr\"om'' type~\eqref{ARN}.

 \item
     $w_0 \ne \pm 1$,
     and the metric function~$u$ is positive in some neighborhood
     of $r = 0$.  In this case, all solutions belong to
     the Reissner--Nordstr\"om type~\eqref{RN3}.
\end{enumerate}
\end{Theorem}

     We remark that this classification scheme explains why
     almost all interior black hole solutions, found numerically
     in~\cite{grqc96}, exhibit oscillatory behavior of the metric.


\section{CRITICAL SETS FOR $r \ne 0$}

\subsection{The critical surface \RH}

     Now let us discuss the remaining critical sets of~\eqref{DS}.
     The surface \RH: $(r, 0, (1 - w^2) r w/[(1 - w^2)^2 - r^2], w)$
     is an unstable hyperbolic set.  The eigenvalues of~\RH{} are
     $\lambda _r = 0$, $\lambda _u = - [(1 - w^2)^2 - r^2]$,
     $\lambda _v = (1 - w^2)^2 -r^2$, and $\lambda _w = 0$.
     The three-dimensional separatrices that are tangent to
     the eigenvector $\zeta_v = (0,0,1,0)$ read as
\[
     r \equiv r_0 = \const{},                             \quad
     u \equiv 0,                                          \quad
     v = \frac{\gamma}{\beta} + C_v \, e^{\beta t},       \quad
     w \equiv w_0 = \const{},
\]
     where $\beta = (1 - w_0^2)^2 - r_0^2 \ne 0$ and
     $\gamma = (1 - w_0^2) r_0 w_0$. Obviously, they do not correspond
     to any EYM solution.  In their turn, the three-dimensional
     separatrices that are tangent to the eigenvector
\[
     \zeta _u = \left(1, -\frac{G}{r},
                \frac{\{2 F^4 - [F^3 + (1 - 3 w^2) r^2] r^2 \} F w}
                     {2 G^3}, \frac{F r w}{G} \right),
\]
     where $F = 1 - w^2$ and $G = F^2 - r^2$, correspond to
     a two-parameter family of local EYM solutions.  It is convenient
     to fix one of these parameters and to write down these solutions
     as follows.

\begin{Proposition}
     For any fixed $r_h > 0$, the EYM equations~\eqref{EYM} possess
     a one-parameter family of local solutions of the form
\begin{equation} \label{as4RegHor}
 \begin{split}
     u & = -\frac{(1 - w_h^2)^2 - r_h^2}{r_h} \, s
              + o(s), \\[2mm]
     w & = w_h + \frac{(1 - w_h^2) r_h w_h}
                       {(1 - w_h^2)^2 - r_h^2} \, s \\[2mm]
       & \quad \, + \frac{(1-w_h^2)
                 \{2 (1 - w_h^2)^4 - [(1 - w_h^2)^3
                   + (1 - 3 w_h^2) r_h^2] r_h^2 \} w_h}
                  {4 [(1 - w_h^2)^2 - r_h^2]^3} \, s^2
       + o(s^2)  \\[2mm]
 \end{split}
\end{equation}
     as $s = r - r_h \to 0$,
     where $w_h$ is a constant, satisfying $\abs{1 - w_h^2} \ne r_h$.
\end{Proposition}

     The local solutions~\eqref{as4RegHor} represent
     the EYM solutions in the vicinity of a regular horizon.
     For the black hole solutions, this is either an event
     horizon (if $r_h > \abs{1 - w_h^2}$), or an interior
     Cauchy horizon (if $r_h < \abs{1 - w_h^2}$).  The first
     existence proof for these local solutions was given
     in~\cite{SWY:bh}.  Some black hole solutions with
     an interior horizon were found numerically in~\cite{grqc96}.

     Note that \RH{} transforms to the line~$W$ for $r=0$.


\subsection{The critical lines \DegH}

     The lines \DegH: $(\pm1, 0, v, 0)$ are degenerate.  All their
     eigenvalues are equal to zero.  Since the EYM equations~\eqref{EYM}
     are invariant under the transformation $r \to -r$, we shall
     study~\eqref{DS} only in the vicinity of the line~$\mathit{DH}^+$.

     It is convenient to use the local coordinates
\[
     \bar{r} = r - 1,                        \quad
     u_r = \frac{u}{\bar{r}^2},              \quad
     v,                                      \quad
     w_r = \frac{w}{\bar{r}},
\]
     in which the dynamical system~\eqref{DS} reads as
\begin{equation} \label{ds4MrrDH}
 \begin{split}
     \dot{\bar{r}} & = (1+ \bar{r}) \, \bar{r} u_r, \\
     \dot{u}_{r}   & = \left[2
                       -\left(2 + \bar{r}+ 2 \bar{r} v^2 \right) u_r
                     +\left(1 + 2 w_r^2 \right) \bar{r}
                       - \bar{r}^3 w_r^4 \right] u_r, \\
     \dot{v}       & = - 2 v - w_r
                       -\left(v + w_r - u_r v
                              + 2 v w_r^2 \right) \bar{r}
                       + \bar{r}^2 w_r^3
                       + (1 + v w_r) \bar{r}^3 w_r^3, \\
     \dot{w}_r     & = (1 + \bar{r}) (v - w_r) \, u_r,
 \end{split}
\end{equation}
     where an overdot stands for derivatives with respect to~$\tau$
     defined by $d\tau = \bar{r} \, dt$.

     There are two critical sets of~\eqref{ds4MrrDH} in the
     hyperplane $\bar{r} = 0$, namely, the point $D$: $(0,1,0,0)$
     and the line $L$: $(0, 0, -\frac{1}{2} w_r, w_r)$.
     The eigenvalues of~$D$ are
     $\lambda _{\bar{r}} = 1$,
     $\lambda _{u_r}     = - 2$, and
     $\lambda _{v, w_r}  = - \tfrac{1}{2} (3 \pm i \sqrt{3})$.
     Thus, $D$ is a saddle.  The outgoing one-dimensional separatrices
\[
     \bar{r} = - \frac{1}{1 + C_r \, e^{-\tau}},         \quad
     u_r     \equiv 1,                                   \quad
     v       \equiv 0,                                   \quad
     w_r \equiv 0,
\]
     which are tangent to the eigenvector $\zeta_{\bar{r}} = (1,0,0,0)$,
     correspond to the extreme Reissner--Nordstr\"om solution
\begin{equation} \label{ERN}
     w  \equiv 0,                                        \quad
     u  = (1 - r)^2.
\end{equation}
     Besides this, in the vicinity of~$D$ there exists a
     stable three-dimensional manifold.
     In order to figure out whether this manifold belongs to
     the hyperplane $\bar{r} = 0$, one may study a projection
     of~\eqref{ds4MrrDH} into $\bar{r} = \const$.  It occurs that
     the critical point $(1,0,0)$ exists for any~$\bar{r}$ and has
     the eigenvalues $\lambda_{u_r} = -2 - \bar{r}$ and
     $\lambda _{v,w_r} = -\frac{1}{2} \, (3 + \bar{r})
     \pm i \sqrt{(3 + \bar{r})(1 + 3 \bar{r})}$.  Hence, for any
     $\bar{r} > -1/3$ all the eigenvalues have negative real part,
     and~$\lambda_v$ and~$\lambda_{w_r}$ are complex conjugate.
     Thus, the stable three-dimensional manifold does not belong
     to the hyperplane $\bar{r} = 0$, and the trajectories on this
     manifold correspond to a two-parameter family of local EYM
     solutions.

     It is interesting to note that the projection of~\eqref{ds4MrrDH}
     into the plane $(\bar{r} = 0, u_r = 1)$ gives a system of two
     linear differential equations
\begin{equation} \label{2Dds4E}
 \begin{split}
     \dot{v}   & = - 2 v - w_r, \\
     \dot{w}_r & = v - w_r,
 \end{split}
\end{equation}
     which can be easily solved:
\[
 \begin{split}
     v   & = \left[C_1 \cos \left( \tfrac{\sqrt{3}}{2} \tau \right)
                 - \tfrac{\sqrt{3}}{3} (C_1 + 2 C_2)
                   \sin \left( \tfrac{\sqrt{3}}{2} \tau \right)
             \right] \, \exp \left(-\tfrac{3}{2} \tau \right), \\
     w_r & = \left[C_2 \cos \left( \tfrac{\sqrt{3}}{2} \tau \right)
                 + \tfrac{\sqrt{3}}{3} (2 C_1 + C_2)
                   \sin \left( \tfrac{\sqrt{3}}{2} \tau \right)
             \right] \, \exp \left(-\tfrac{3}{2} \tau \right),
 \end{split}
\]
     where $C_1$ and $C_2$ are the constants of integration.
     Thus, the projection of~\eqref{ds4MrrDH}  into
     $(\bar{r} = 0, u_r = 1)$ represents linear oscillations
     of~$v$ and~$w_r$ with infinitely many zeros.

     It is easy to see that $\bar{r} \to 0$ corresponds to
     $\tau \to -\infty$, so that~$w'$ diverges as $r \to 1$.
     Conversely,~$r$ goes away from~$1$ as $\tau \to +\infty$,
     and these solutions tend to the extreme Reissner--Nordstr\"om
     solution.  Thus, we have
\begin{Proposition}
     There exists a neighborhood of $r=1$, in which the EYM
     equations~\eqref{EYM} have a two-parameter family of local
     solutions with the gauge function oscillating with infinitely
     many zeros, and the metric function tending to zero.
\end{Proposition}

     These solutions can be treated as a description
     of the limiting behavior of the EYM solutions in the vicinity
     of $r = 1$ as the number of the gauge function nodes tends to
     infinity.  The limiting behavior of the EYM solutions was first
     studied in~\cite{KMuA,SW:LimBeh,SW:LimMass}.  Solutions that
     exhibit oscillations of~$w$ were first discussed in~\cite{BFM94}.
     But, in addition to the results of~\cite{BFM94}, we see that
     the gauge function may have infinitely many zeros not only to
     the left of $r=1$, but also to the right.

\medskip

     Finally, the line~$L$ has the eigenvalues $\lambda_{\bar{r}} = 0$,
     $\lambda_{u_r} = 2$, $\lambda_v = -2$, and $\lambda_{w_r} = 0$.
     Thus, $L$ is a degenerate set.  The eigenvectors
     $\zeta_{u_r} = (0,1,\frac{3}{16} w_r, -\frac{3}{4} w_r)$ and
     $\zeta_v = (0,0,1,0)$ determine two-dimensional separatrices,
     which lie in the hyperplane $\bar{r} = 0$.  Thus, they do not
     correspond to any EYM solution.  Further investigation of the
     line~$L$ did not reveal any trajectories of~\eqref{DS} that have
     corresponding EYM solutions different from the discussed above.


\section{SOLUTIONS WITH A SINGULAR HORIZON}

     A typical EYM solution cannot be continued to the far field, since
     it has a singular horizon, i.e., a point, at which the metric
     function tends to zero, the gauge functions stays finite, but
     its derivative diverges~\cite{SW:particle,SWY:bh}.  This fact was
     first noticed in~\cite{KMuA}, and a power series expansion,
     describing the behavior of the EYM solutions in the vicinity of a
     singular horizon was given.  Let us show how one can obtain
     singular EYM solutions basing on dynamical systems methods.
     This will also lead us to a discovery of two new local solutions.

     Let us rewrite the dynamical system~\eqref{DS} as
\begin{equation} \label{ds4z=1/v}
 \begin{split}
     \dot{r} & = r u z^2, \\
     \dot{u} & = -\left[\left(2 - z^2 \right) u
                 + \left(1 - w^2 \right)^2 z^2
                 - r^2 z^2 \right] u, \\
     \dot{z} & = -\left[u + \left(1 - w^2 \right)^2 - r^2
                 - \left(1 - w^2 \right) r z w \right] z^3, \\
     \dot{w} & = r u z,
 \end{split}
\end{equation}
     where $z = 1/v$, and an overdot stands for derivatives with
     respect to~$\tau$ defined by $dt = z^2 \, d\tau$.

     For $z = 0$, the critical points of~\eqref{ds4z=1/v} form a
     degenerate plane $(r, 0, 0, w)$.  In order to
     study~\eqref{ds4z=1/v} in the vicinity of this plane, we
     introduce the local coordinates $(r, u_z = u/z^2, z, w)$,
     in which~\eqref{ds4z=1/v} can be written as
\begin{equation} \label{ds4SingHrzn}
 \begin{split}
     \dot{r}   & = r u_z z^2, \\
     \dot{u}_z & = - \left[\left(2 - 3 z^2 \right) u_z
                   - \left(1 - w^2 \right)^2 + r^2
                   + 2 \left(1 - w^2 \right) r z w \right] u_z, \\
     \dot{z}   & = - \left[u_z z^2 + \left(1 - w^2 \right)^2 - r^2
                      - \left(1 - w^2 \right) r z w \right] z, \\
     \dot{w}   & = r u_z z,
 \end{split}
\end{equation}
     where an overdot stands for derivatives with respect to~$t$.

     The system~\eqref{ds4SingHrzn} has two critical sets in the
     hyperplane $z = 0$, namely, the plane~$RW_1$: $(r, 0, 0, w)$ and
     the surface~$\mathit{SH}$: $(r, [(1 - w^2)^2 - r^2]/2, 0, w)$,
     which are nondegenerate whenever
\begin{equation} \label{NonDeg}
                          r \ne \abs{1 - w^2}.
\end{equation}

     The eigenvalues of~$RW_1$ are $\lambda _r = 0$,
     $\lambda _{u_z} = (1 - w^2)^2 - r^2$,
     $\lambda _z  = - \lambda _{u_z}$, and $\lambda _w = 0$.
     Thus, $RW_1$ is an unstable hyperbolic set.
     One can easily see that if~\eqref{NonDeg} holds, then
     the eigenvectors $\zeta_{u_z} = (0,1,0,0)$ and
     $\zeta_z = (0,0,1,0)$ determine the three-dimensional separatrices
\[
     r   \equiv r_0 = \const{},                             \quad
     u_z = \frac{\beta}{2 + C_u \, e^{-\beta t}},           \quad
     z   \equiv 0,                                          \quad
     w   \equiv w_0 = \const{},
\]
     and
\[
     r   \equiv r_0,                                        \quad
     u_z \equiv 0,                                          \quad
     z   = \frac{\beta}{\gamma + C_z \, e^{\beta t}},       \quad
     w   \equiv w_0,
\]
     respectively, where $\beta = (1 - w_0^2)^2 -  r_0^2 \ne 0$
     and $\gamma = (1 - w_0^2) r_0 w_0$.  Clearly, these
     separatrices do not correspond to any EYM solution.

     Next, the eigenvalues of~$\mathit{SH}$ are
     $\lambda _r     = 0$,
     $\lambda _{u_z} = \lambda _z = - [(1 - w^2)^2 - r^2]$, and
     $\lambda _w     = 0$.
     Hence, if~\eqref{NonDeg} holds, then all trajectories
     of~\eqref{ds4SingHrzn} in the vicinity of~$\mathit{SH}$
     belong to a four-dimensional manifold.
     The eigenvalues~$\lambda _{u_z}$ and~$\lambda _z$ have the
     eigenvectors $\zeta_{u_z} = (0,1,0,0)$ and
     $\zeta_z = (0,0,1,-\frac{1}{2} r)$.  Thus, all trajectories
     in the vicinity of~$\mathit{SH}$ take the form
\[
     r   = r_0 + o(z),                              \quad
     u_z = \frac{\beta}{2} + u_1 z + o(z),          \quad
     w   = w_0 - \frac{r_0}{2} z + o(z)             \quad
\]
     as $z \to 0$, where $\beta \ne 0$, and $u_1$ is an arbitrary
     constant.  This leads to

\begin{Proposition}
     Let $r_0 > 0$ be a point such that
\[
     \lim _{r \to r_0} u(r) = 0,                       \quad
     \lim _{r \to r_0} w(r) = w_0 < \infty ,           \quad
     \lim _{r \to r_0} w'(r) = \infty,
\]
     and $\beta = (1 - w_0^2)^2 -  r_0^2 \ne 0$.  Then
     all solutions of the EYM equations~\eqref{EYM}
     in the vicinity of~$r_0$ have the form
\begin{equation} \label{as4SinHor}
     u = \frac{2 \beta}{r_0} s^2 + u_1 s^3 + o(s^3), \quad
     w = w_0 \pm \sqrt{r_0} \, s + o(s)
\end{equation}
     as $s = \sqrt{r_0 - r} \to 0$,
     where $u_1$ is an arbitrary constant.  These solutions do
     not have other parameters besides~$w_0$ and~$u_1$.
\end{Proposition}

     Thus, local solutions~\eqref{as4SinHor} exist in the vicinity of
     any point $r_0 > 0$, $r_0 \ne \abs{1 - w_0^2}$.

     Notice that solutions~\eqref{as4RegHor} were determined by
     the trajectories that belong to a three-dimensional manifold.
     Unlike them, singular solutions~\eqref{as4SinHor} correspond
     to the trajectories that lie on a four-dimensional manifold.
     It follows immediately that in the vicinity of an arbitrary point
     $r_0 > 0$ such that $\lim _{r \to r_0} u(r) = 0$,
     $\lim _{r \to r_0} w(r) = w_0$, and  $r_0 \ne \abs{1 - w_0^2}$,
     almost all solutions of the EYM equations~\eqref{EYM} exhibit
     asymptotic behavior~\eqref{as4SinHor} and, therefore, cannot be
     continued toward $r = \infty$.

     Let us also mention that it follows from~\eqref{as4SinHor} and the
     above analysis of the critical sets~\RH{} and~\DegH{} that almost
     all EYM solutions, defined at an arbitrary finite point $r_0 > 0$,
     can be continued to the left for all $r < r_0$.  A detailed
     investigation of extendibility of solutions of the EYM equations
     can be found in~\cite{SW:extend}.


\medskip

     Now let us study~\eqref{ds4z=1/v} in the vicinity of the
     curve $r = \abs{1 - w^2}$.  We start with $\abs{w} \le 1$.
     In the local coordinates
\[
     r_z = \frac{1}{z} \, (r - 1 + w^2),               \quad
     u_z = \frac{u}{z^3},                              \quad
     z,                                                \quad
     w,
\]
     the dynamical system~\eqref{ds4z=1/v} reads as
\begin{equation} \label{ds4Rz}
 \begin{split}
     \dot{r}_z & = \left(1 - w^2 \right) \bigl(z + 2 w \bigr) u_z
                    + \left[2 (z + w) u_z z
                         - \left(1 - w^2 \right)^2 w \right] r_z    \\
         & \quad \, - \left(1 - w^2 \right) \bigl(2 + z w \bigr) r_z^2
                    - r_z^3 z,                                       \\
     \dot{u}_z & = - \left[2 \left(1 - 2 z^2 \right) u_z
                    + 3 \left(1 - w^2 \right)^2 w
                    + \left(1 - w^2 \right) \bigl(4 + 3 z w \bigr) r_z
                    + 2 r_z^2 z \right] u_z,                         \\
     \dot{z}   & = \left[ \left(1 - w^2 \right)^2 w - u_z z^2
                    + \left(1 - w^2 \right) \bigl(2 + z w \bigr) r_z
                    + r_z^2 z \right] z,                             \\
     \dot{w}   & = \left(1 - w^2 + r_z z \right) u_z z,
 \end{split}
\end{equation}
     where an overdot stands for derivatives with respect to~$\tau$
     defined by $d\tau = z \, dt$.

     The system~\eqref{ds4Rz} has six critical sets in the hyperplane
     $z = 0$, namely, \LW: $(0, 0, 0, w)$, \LR: $(r_z, 0, 0, \pm 1)$,
     \CR: $(-\tfrac{1}{2}(1 - w^2) w, 0, 0, w)$, $RU_1$:
     $(-(1 - w^2) w, \tfrac{1}{2}(1 - w^2)^2 w, 0, w)$, and $RU_2$:
     $(-\tfrac{3}{2}(1 - w^2) w, \tfrac{3}{2}(1 - w^2)^2 w, 0, w)$.
     Analysis of the first four critical sets does not reveal any
     trajectories of~\eqref{ds4Rz} that have corresponding
     EYM solutions.
     Thus we discuss only the curves~$RU_1$ and~$RU_2$.

     The eigenvalues of $RU_1$ are
     $\lambda _{r_z} = \lambda _{u_z} = (1 - w^2)^2 w$,
     $\lambda _{z}   = - (1 - w^2)^2 w$, and
 $\lambda _w     = 0$.
     Thus, $RU_1$ is an unstable hyperbolic critical set whenever
\begin{equation} \label{wne01}
     w \ne 0, \pm 1.
\end{equation}
     In this case, all trajectories on a three-dimensional manifold,
     determined by~$\lambda _{r_z}$ and~$\lambda_{u_z}$, are tangent to
     the eigenvector $\zeta_{r_z,u_z} = (1, -(1 - w^2),0,0)$, which
     defines the two-dimensional separatrices
\begin{equation*}
     u_z = - \tfrac{1}{2} \alpha^2 w_0 - \alpha r_z,        \quad
     z   \equiv 0,                                          \quad
     w   \equiv w_0 \ne 0, \pm 1.
\end{equation*}
     One of these separatrices joins~$RU_1$ to~\CR.
     However, the hole manifold belongs to the hyperplane $z = 0$.
     Thus, the trajectories on it do not correspond to any EYM solution.

     Unlike this, the two-dimensional separatrices that are tangent to
     the eigenvector
\[
     \zeta _z = \left(
      \tfrac{1}{8} \, \left(1 - w^2 \right) \left(3 - 5 w^2 \right),
     -\tfrac{1}{4} \, \left(1 - w^2 \right)^2 \left(2 - 3 w^2 \right),
      1,
     -\tfrac{1}{2} \, \left(1 - w^2 \right) \right),
\]
     take the form
\begin{equation}  \label{as4RU1}
 \begin{split}
     r_z & = - \alpha w_0
             + \tfrac{1}{8} \, \alpha \left(3 - 5 w_0^2 \right) z
             + o(z),                                             \\
     u_z & = \tfrac{1}{2} \, \alpha^2 w_0
             - \tfrac{1}{4} \, \alpha^2 \left(2 - 3 w_0^2 \right) z
             + o(z),                                             \\
     w   & = w_0  - \tfrac{1}{2} \, \alpha z + o(z)
 \end{split}
\end{equation}
     as $z \to 0$,
     where $w_0 \ne 0, \pm1$, and thus have corresponding EYM
     solutions.

     Finally, the eigenvalues of~$RU_2$ are
     $\lambda _{r_z} = -   (1 - w^2)^2 w$,
     $\lambda _{u_z} =   3 (1 - w^2)^2 w$,
     $\lambda _{z}   = - 2 (1 - w^2)^2 w$, and $\lambda _w     = 0$.
     Hence, $RU_2$ is also an unstable hyperbolic critical set
     whenever~\eqref{wne01} holds.  The eigenvector
     $\zeta _{u_z} = (1, -(1 - w^2), 0, 0)$ defines the two-dimensional
     separatrices
\begin{equation*}
     u_z = - \alpha r_z,                                   \quad
     z   \equiv 0,                                         \quad
     w   \equiv w_0 \ne 0, \pm 1.
\end{equation*}
     One of them joins~$RU_2$ to~\LW.  Besides these separatrices,
     there is also a three-dimensional manifold, defined by the
     eigenvalues~$\lambda _{r_z}$ and~$\lambda_z$.  Almost all
     trajectories on this manifold are tangent to the eigenvector
     $\zeta_{r_z} = (1, -3 (1 - w^2),0,0)$, which determines
     the separatrices
\[
     u_z  = \tfrac{3}{2} \, \alpha^2 w_0
                  - 3 \alpha s + o(s)                       \quad
     \text{as }
     s    = r_z  + \tfrac{3}{2} \, \alpha w_0 \to 0,        \qquad
     z    \equiv 0,                                         \quad
     w    \equiv w_0,
\]
     where $w_0 \ne 0, \pm 1$. In addition, there exist two-dimensional
     separatrices,  which are tangent to the eigenvector
\[
     \zeta _z = \left(
      \tfrac{3}{40} \, \left(1 - w^2 \right) \left(13 - 12 w^2 \right),
     -\tfrac{9}{40} \, \left(1 - w^2 \right)^2 \left(11 - 9 w^2 \right),
      1,
     -\tfrac{3}{4} \, \left(1 - w^2 \right) \right)
\]
     and take the form
\begin{equation}  \label{as4RU2}
 \begin{split}
     r_z & = - \tfrac{3}{2} \,\alpha w_0
             + \tfrac{3}{40} \, \alpha \left(13 - 12 w_0^2 \right) z
             + o(z),                                             \\
     u_z & = \tfrac{3}{2} \, \alpha^2 w_0
             - \tfrac{9}{40} \, \alpha^2 \left(11 - 9 w_0^2 \right) z
             + o(z),                                             \\
     w   & = w_0  - \tfrac{3}{4} \, \alpha z + o(z)
 \end{split}
\end{equation}
     as  $z \to 0$, where $w_0 \ne 0, \pm 1$.  These separatrices,
     together with~\eqref{as4RU1}, have corresponding singular EYM
     solutions. Conversely, all trajectories, defined
     by the eigenvalues~$\lambda_{u_z}$ and~$\lambda_{r_z}$,
     belong to the hyperplane $z = 0$ and do not have counterparts
     neither for the dynamical system~\eqref{ds4z=1/v}, nor
     for the EYM equations.

     Analysis of~\eqref{ds4z=1/v} in the vicinity of the curve
     $r = -(1 - w^2)$ for $\abs{w} \ge 1$ is completely analogous
     to the previous case.  The dynamical system~\eqref{ds4z=1/v},
     written in the local coordinates
\[
     r_z = \frac{1}{z} \, (r + 1 - w^2),               \quad
     u_z,                                              \quad
     z,                                                \quad
     w,
\]
     has the same critical sets in the hyperplane $z = 0$,
     as~\eqref{ds4Rz}, to the exclusion of the
     curves~$RU_1$ and~$RU_2$, which in this case have the opposite
     sign of~$u_z$.  Asymptotic formulas~\eqref{as4RU1}
     and~\eqref{as4RU2} convert to
\begin{align*}
     r_z & = - \alpha w_0
             - \tfrac{1}{8} \, \alpha \left(3 - 5 w_0^2 \right) z
             + o(z),                                             \\
     u_z & = - \tfrac{1}{2} \, \alpha^2 w_0
             - \tfrac{1}{4} \, \alpha^2 \left(2 - 3 w_0^2 \right) z
             + o(z),                                             \\
     w   & = w_0  + \tfrac{1}{2} \, \alpha z + o(z),
\intertext{and}
     r_z & = - \tfrac{3}{2} \,\alpha w_0
             - \tfrac{3}{40} \, \alpha \left(13 - 12 w_0^2 \right) z
             + o(z),                                             \\
     u_z & = - \tfrac{3}{2} \, \alpha^2 w_0
             - \tfrac{9}{40} \, \alpha^2 \left(11 - 9 w_0^2 \right) z
             + o(z),                                             \\
     w   & = w_0  + \tfrac{3}{4} \, \alpha z + o(z)
\end{align*}
     as $z \to 0$,
     respectively.  Recall that $\alpha = 1 - w_0^2 = - r_0 \ne 0,1$
     here.  Combining these solutions with~\eqref{as4RU1}
     and~\eqref{as4RU2}, we get

\begin{Proposition} \label{prop4parabola}
     In the vicinity of any point $r_0 > 0$, $r_0 \ne 1$,
     the EYM equations~\eqref{EYM} have solutions of the form
\[
     u = \pm 4 \xi \sqrt{r_0} \, w_0 \, s^3 + O(s^4), \quad
     w = w_0 \pm \sqrt{r_0} \, s + o(s),
\]
     and
\[
     u = \pm \frac{16}{3} \, \xi w_0 w_{12} \, s^3 + O(s^4), \quad
     w = w_0 \pm w_{12} \, s + o(s),
\]
     as $s = \sqrt{r_0 - r} \to 0$,
     where $r_0 = \abs{1 - w_0^2}$, $\xi = - \sign (1 - w_0^2)$, and
     $w_{12} = \sqrt{3 r_0/2}$.  These solutions do not have free
     parameters.
\end{Proposition}

     To the best of the author's knowledge, the presented local
     solutions are new.

     Notice that Proposition~\ref{prop4parabola}
     excludes the cases $r_0 = 1$ and
     $r_0 = 0$, which correspond to $w_0 = 0, \pm 1$.  Analysis of
     the first one reveals complex solutions of the EYM equations.
     We do not discuss them here, since their physical interpretation
     is unclear.  The latter case has already been discussed in
     Sec.~\ref{section:r=0}.


\section{SOLUTIONS IN THE FAR FIELD}

     The behavior of the EYM solutions in the far field, $r \gg 1$, was
     studied in great details, see~\cite{SW:RegSolns} and references
     therein.  In this section, we briefly give another existence proof
     for the asymptotically flat solutions and obtain a description
     of the limiting behavior of the EYM solutions as the number of the
     gauge function nodes tends to infinity.  To implement this task,
     we return to the EYM equations~\eqref{eym4Nr},
     but we change~$r$ to $z = 1/r$.  Next,
     we rewrite~\eqref{eym4Nr} as a dynamical system of the form
\begin{equation} \label{ds4N}
 \begin{split}
     \dot{z} & = z^2 N, \\
     \dot{N} & = \left[-1 + \left(1 + 2 z^4 v^2 \right) N
                 + \left(1 - w^2 \right)^2 z^2 \right] z N, \\
     \dot{v} & = \left[1 - 3 N
                 - \left(1 - w^2 \right)^2 z^2 \right] z v
                 - \left(1 - w^2 \right) w, \\
     \dot{w} & = z^2 v N,
 \end{split}
\end{equation}
     where $v = w'(z)$, and an overdot stands for derivatives
     with respect to ~$t$ defined by $dz =  z^2 N \, dt$.

     The dynamical system~\eqref{ds4N} has two critical sets
     in the hyperplane $z = 0$, namely, the planes
     \AF: $(0, N, v, \pm 1)$ and \OS: $(0, N, v, 0)$.  Both of
     them are degenerate.  Thus we perform the standard procedure
     of their investigation for finite~$N$ and~$v$.


\subsection{The critical planes \AF}

     In this case, we introduce the local coordinates
     $(z, N, v, w_z = \bar{w}/z)$,
     where $w = \bar{w} \pm 1$; here the upper sign applies
     for~$\mathit{AF}^+$ and the lower one for~$\mathit{AF}^-$.
     Now~\eqref{ds4N} can be written as
\begin{equation} \label{ds4MzAF}
 \begin{split}
     \dot{z}   & = z N, \\
     \dot{N}   & = \left[-1 + \left(1 + 2 z^4 v^2 \right) N
                    + (2 \pm z w_z)^2 z^4 w_z^2 \right] N, \\
     \dot{v}   & = \left[1 - 3 N
                    - (2 \pm z w_z)^2 z^4 w_z^2 \right] v
                    + (1 \pm z w_z)(2 \pm z w_z) w_z, \\
     \dot{w}_z & = (v - w_z) N,
 \end{split}
\end{equation}
     where an overdot stands for derivatives with respect to~$\tau$
     defined by $d\tau = z \, dt$.

     The dynamical system \eqref{ds4MzAF} has two critical lines in
     the hyperplane $z = 0$, namely, $Z_1$: $(0, 1, w_z, w_z)$ and
     $Z_2$: $(0, 0, -2 w_z, w_z)$.  The eigenvalues of~$Z_1$ are
     $\lambda _{z}   = \lambda _{N} = 1$,
     $\lambda _v     = -3$, and
 $\lambda _{w_z} = 0$.
     The eigenvector $\zeta_v = (0, 0, -2, 1)$ defines the
     ingoing two-dimensional separatrices
\[
     z \equiv 0,                     \quad
     N \equiv 1,                     \quad
     w_z = w_0 - \tfrac{1}{2} v,
\]
     where $w_0$ is an arbitrary constant. Since these separatrices
     belong to the hyperplane $z = 0$, they do not correspond to any
     trajectories of~\eqref{ds4N}.

     The eigenvalues~$\lambda _{z}$ and~$\lambda _{N}$ have the
     eigenvectors
     $\zeta _z = \left(1, 0, \pm \tfrac{3}{2} w_z^2,
       \pm \tfrac{3}{4} w_z^2 \right)$ and
     $\zeta _N = \left(0, 1, -\tfrac{3}{2} w_z, -\tfrac{3}{4} w_z
       \right)$,
     where the signs in~$\zeta _z$ correspond to
     the signs in the right hand sides of~\eqref{ds4MzAF}.
     Thus, the three-dimensional outgoing separatrices take the form
\[
 \begin{split}
     N    & = 1 + n_1 z + o(z),                                    \\
     v    & = w_1 + \tfrac{3}{2} (\pm w_1 - n_1) w_1 z + o(z),     \\
     w_z  & = w_1 + \tfrac{3}{4} (\pm w_1 - n_1) w_1 z + o(z)
 \end{split}
\]
     as $z \to 0$,
     where $n_1$ and $w_1$ are arbitrary constants.  Clearly, these
     separatrices correspond to a two-parameter family of the
     asymptotically flat solutions of~\eqref{eym4Nr}.

\begin{Proposition}
     The EYM equations~\eqref{eym4Nr} possess a two-parameter family of
     solutions such that $\lim_{r \to \infty} w(r) = w_\infty =\pm 1$.
     All these solutions have the form
\[
 \begin{split}
     N & = 1 + n_{-1} r^{-1} + o(r^{-1}), \\
     w & = w_\infty + w_{-1} r^{-1}
           + \tfrac{3}{4}
             (w_\infty w_{-1} - n_{-1}) w_{-1} r^{-2} + o(r^{-2})
 \end{split}
\]
     as $r \to \infty$,
     where $n_{-1}$ and $w_{-1}$ are arbitrary constants.
\end{Proposition}

\medskip

     Next, the eigenvalues of~$Z_2$ are
     $\lambda _z = 0$,   $\lambda _N = -1$,  $\lambda _v = 1$, and
     $\lambda _{w_z} = 0$.
     Hence, $Z_2$ is an unstable degenerate set.  The eigenvalues
     $\zeta_N = (0,1,-6 w_z, 3 w_z)$ and $\zeta_v = (0,0,1,0)$
     determine two-dimensional separatrices, which belong to the
     hyperplane $z = 0$.  Thus, they have no corresponding
     trajectories of~\eqref{ds4N}.  Closer analysis of~$Z_2$ did
     not reveal
     any trajectories of~\eqref{ds4MzAF} that correspond to EYM
     solutions.


\subsection{The critical plane \OS}

     In this case, we study~\eqref{ds4N} in the local coordinates
     $(z, N, v, w_z = w/z)$, in which~\eqref{ds4N} may be written as
\begin{equation} \label{ds4MzOS}
 \begin{split}
     \dot{z}   & = z N, \\
     \dot{N}   & = \bigl[-1 + \left(1 + 2 z^4 v^2 \right) N
                    + \left(1 - z^2 w_z^2 \right)^2 z^2 \bigr] N, \\
     \dot{v}   & = \bigl[1 - 3 N
                    - \left(1 - z^2 w_z^2 \right)^2 z^2 \bigr] v
                    - \left(1 - z^2 w_z^2 \right) w_z, \\
     \dot{w}_z & = (v - w_z) N,
 \end{split}
\end{equation}
     where an overdot stands for derivatives with respect
     to~$\tau$.

     The system~\eqref{ds4MzOS} has two critical sets in the hyperplane
     $z = 0$, namely, the point $P$: $(0,1,0,0)$ and the line $Z_3$:
     $(0, 0, w_z, w_z)$.  The eigenvalues of~$P$ are
     $\lambda _{z}     = \lambda _{N} = 1$ and
     $\lambda _{v,w_z} = - \tfrac{1}{2} (3 \pm i \sqrt{3})$.
     Thus, $P$ is a saddle. The eigenvectors $\zeta_z = (1,0,0,0)$ and
     $\zeta_N = (0,1,0,0)$ determine the outgoing two-dimensional
     separatrices
\[
     N   = 1 + n_1 z + z^2,                        \quad
     v   \equiv 0,                                 \quad
     w_z \equiv 0,
\]
     where $n_1$ is an arbitrary constant.  Obviously, these
     separatrices correspond to the Reissner--Nordstr\"om
     solution~\eqref{ExactRN}.

     Next, trajectories that belong to a stable two-dimensional
     manifold defined by the eigenvalues~$\lambda_v$
     and~$\lambda_{w_z}$, spiral toward~$P$ as $\tau \to +\infty$.
     These solutions may be written down explicitly, since for
     $z \equiv 0$ and $N \equiv 1$ the system~\eqref{ds4MzOS}
     reads exactly as~\eqref{2Dds4E} with~$w_r$ replaced by~$w_z$.
     However, the whole manifold belongs to the hyperplane $z = 0$,
     so that the trajectories on it do not correspond to any EYM
     solution.  One may treat these trajectories as a description of
     the limiting behavior of the EYM solutions as the number of the
     gauge function nodes tends to infinity.

     Finally, the line~$Z_3$ is an unstable degenerate critical set.
     The eigenvalues of~$Z_3$ are $\lambda_z = 0$,  $\lambda_N = -1$,
     $\lambda_v = 1$, and $\lambda_{w_z} = 0$. One can easily see that
     the eigenvectors $\zeta_N = (0, 1, \frac{3}{2} w_z, 0)$ and
     $\zeta_v = (0,0,1,0)$ define two-dimensional separatrices, which
     belong to the hyperplane $z = 0$.  Thus, they do not have
     corresponding trajectories of~\eqref{ds4N}. Additional study
     of~$Z_3$ did not reveal any trajectories of~\eqref{ds4MzOS}
     that have corresponding EYM solutions.

\medskip

     Let us mention in conclusion that though our investigation
     was restricted to local solutions of the EYM equations,
     dynamical systems methods can also be used for the analysis
     of the solutions global behavior.  This will be the subject
     of a forthcoming publication.


\section*{Acknowledgments}
     The author thanks Prof.\ D.~V.~Gal'tsov for suggesting the
     problem and for constant attention to the investigation, Prof.\
     O.~I.~Bogoyavlensky for explaining some details of the method
     used in~\cite{OB} and for comments on the manuscript, Profs.\
     Yu.~A.~Fomin and G.~V.~Kulikov for their sustained support,
     and Prof.\ J.~A.~Smoller for kindly sending numerous articles
     on the EYM equations.

     The work was partially supported by the RFBR, Grant No.~96-02-18899.



\end{document}